# Room Temperature All Semiconducting sub-10nm Graphene Nanoribbon Field-Effect Transistors

Xinran Wang,[1] Yijian Ouyang,[2] Xiaolin Li,[1] Hailiang Wang,[1] Jing Guo,[2] and Hongjie Dai[1,*]

[1] *Department of Chemistry and Laboratory for Advanced Materials, Stanford University, Stanford, CA 94305, USA*

[2] *Department of Electrical and Computer Engineering, University of Florida, Gainesville, FL, 32611*

**Abstract**

Sub-10nm wide graphene nanoribbon field-effect transistors (GNRFETs) are studied systematically. All sub-10nm GNRs afforded semiconducting FETs without exception, with $I_{on}/I_{off}$ ratio up to $10^6$ and on-state current density as high as ~2000μA/μm. We estimated carrier mobility ~200cm$^2$/Vs and scattering mean free path ~10nm in sub-10nm GNRs. Scattering mechanisms by edges, acoustic phonon and defects are discussed. The sub-10nm GNRFETs are comparable to small diameter ($d$≤~1.2nm) carbon nanotube FETs with Pd contacts in on-state current density and $I_{on}/I_{off}$ ratio, but have the advantage of producing all-semiconducting devices.

* Correspondence should be addressed to hdai@stanford.edu

Graphene-based electronics has attracted much attention due to high carrier mobility in bulk graphene[1-5]. For mainstream logic applications, graphene width confinement down to sub-10nm is needed to open sufficient bandgap for room temperature transistor operation. Although sub-10nm GNR was predicted to be semiconducting by several theories[6-10], experimental work in this area[11,12] has been scarce partly due to challenges in patterning GNR below 20nm by plasma etching. Recently, sub-10nm GNRs with smooth edges were obtained and demonstrated to be semiconductors with bandgap inversely proportional to $w$ (Ref. 13). Various fundamental questions remain to be addressed such as the performance limit of GNRFETs, the intrinsic carrier mobility in narrow ribbons and comparison of GNRs with other materials including carbon nanotubes.

In this work, we studied both sub-10nm GNRs and wide GNRs ($w\sim$10-60nm). All the sub-10nm GNRs (a total of $\sim$ 40) were found semiconducting with adequate bandgap for transistor operation at room temperature. The GNR synthesis and transistor fabrication process (see Supplementary Information) were similar to that described in Ref. 13. Fig. 1b and c show AFM images of typical sub-10nm ($w\sim2\pm0.5$nm) and wide ($w\sim60\pm5$nm) GNR devices. We carefully used AFM to measure the width (with careful tip size correction), lengths and number of layers of our GNR devices. Only a few discrete heights have been observed for all the GNR samples we made, i.e. ~1.1nm, 1.5nm and 1.9nm, which were assigned as 1, 2, and 3-layer graphene[12,13]. All of the devices presented in this paper show a height of ~1.5nm and are assigned as 2-layer GNRs, unless stated otherwise. We also carried out confocal surface enhanced raman spectroscopy study on GNR devices. All the details and results are described in supplementary information.

Since our GNRFETs were Schottky barrier (SB) type FETs where the current was modulated by carrier tunnelling probability through SB at contacts, high work function metal Pd was used to minimize the SB height for holes in p-type transistors. In fact we used Ti/Au as contact and found that Pd did give higher $I_{on}$ in device with similar dimensions. 10nm $SiO_2$ gate dielectrics was also important to achieve higher $I_{on}$ because it significantly reduced SB width at contacts compared to 300nm in previous work[13]. Fig. 2a and 2b showed the transfer and output characteristics for the $w\sim2\pm0.5$nm $L\sim236$nm GNR device shown in Fig. 1b. This device delivered $I_{on} \sim$ 4µA (~2000µA/µm) at $V_{ds}$=1V, $I_{on}/I_{off}$ ratio >$10^6$ at $V_{ds}$=0.5V, subthreshold slope ~210mV/decade and transconductance ~1.8µS (~900µS/µm). For wide GNR devices, they all showed metallic behavior because of vanishingly small bandgaps (Fig. 2c and d). Compared to sub-10nm GNRFETs with similar channel length, the current density in wide GNR devices was usually higher (~3000µA/µm at $V_{ds}$=1V for the device in Fig. 2d). We note that our wide GNRs showed relatively weak gate dependence in transfer characteristics, likely due to interaction between layers[12]. The Dirac point was usually not observed around zero gate bias, indicating p-doping effects at the edges or by physisorbed species during the chemical treatment steps[14].

To investigate the intrinsic properties of GNRs such as carrier scattering mean free path (mfp) and mobility, we made different channel length transistors on the same GNR. Fig. 3a showed an AFM image of a typical $w\sim2.5\pm1$nm GNR with $L\sim$110nm, 216nm and 470nm segments that delivered $I_{on}$ ~5µA, ~4µA and ~2µA, respectively (only the output characteristics of the upper segment is shown in Fig. 3b). We measured the low bias resistance $R_{tot}$ of the three segments (Fig. 3c), and extrapolated the parasitic contact resistance $R_c$=~60kΩ for this device. Under low bias, the on-state resistance in GNR due to scattering can be written as

$$R = R_{tot} - R_c = \left(\frac{h}{2e^2}\right)\left(\frac{L}{\lambda}\right) = L\left(\frac{1}{\lambda_{edge}} + \frac{1}{\lambda_{ap}} + \frac{1}{\lambda_{defect}}\right)\left(\frac{h}{2e^2}\right), \quad (1)$$

where $L$ is channel lengths, $\lambda$ is total scattering mfp and $\lambda_{edge}$, $\lambda_{ap}$, $\lambda_{defect}$ denote mfp due to GNR edge, acoustic phonon and defect scattering, respectively. The scattering mfp of GNR

$$\lambda = L\left(\frac{h}{2e^2}\right) \Big/ (R_{tot} - R_c) \quad (2)$$

We estimated λ~14nm, 11nm and 12nm in the three segments of the GNR. Based on standard transistor model, the intrinsic carrier mobility is

$$\mu = \frac{g_m L}{C_{gs} V_{ds}}, \quad (3)$$

where $g_m = \frac{dI_{ds}}{dV_{gs}}\Big|_{V_{ds}}$ is the intrinsic transconductance obtained from the measured $g_m^{mes}$ by excluding the source resistance $g_m = g_m^{mes}/(1 - g_m^{mes} R_s)$ and $C_{gs}$ is gate capacitance per unit length. We used three-dimensional electrostatic simulation to calculate $C_{gs}$ (see Supplementary Information) and obtained $C_{gs}$~26pF/m for a w~2.5nm ribbon. Using Eq. 3, we calculated μ~174cm$^2$/Vs, 171cm$^2$/Vs and 189cm$^2$/Vs in the three segments after excluding the effects of contact resistance. Figure 3b compares the computed $I_{ds}$ vs. $V_{ds}$ characteristics by using a square law model in series with the parasitic resistance to the experimental data for the 470nm GNRFETs.

In narrow GNRs, edge may play an important role. When electrons travel to an edge, a scattering event happens if the edge is not perfect. The edge scattering mfp is modelled as (see Supplementary Information)

$$\lambda_{edge} = \frac{1}{P}\frac{k_{\parallel}}{k_{\perp}}w = \frac{w}{P}\sqrt{\left(1 + \frac{E_k}{\Delta}\right)^2 - 1}, \quad (4)$$

where $k_{\parallel}$ and $k_{\perp}$ are k-space wave vectors along and perpendicular to the GNR direction, $E_k$ is the kinetic energy of electrons, $\Delta$ is half bandgap energy and P is the probability of backscattering which depends on edge quality. From experiment, our low field $\lambda \approx 12$nm < $\lambda_{edge}$, suggesting a backscattering probability P<20% for this ribbon. Assuming similar edge quality in various widths GNRs, our model predicts that $\lambda_{edge}$ is proportional to $w$. Experimentally, we fabricated multiple channel length GNRFETs with different width ribbons and observed the trend that wider sub-10nm GNRs tent to have higher mobility (Fig. 4a, data obtained from multi-probe measurements excluding contact resistance) and mfp, although there were some device-dependent fluctuations. Acoustic phonon and defect scattering can also be responsible for the short mfp. Although $\lambda_{ap}$~10μm is predicted for a $w$~2.5nm hydrogen terminated zigzag GNR[15], we expect it shorter in our GNRs since the edge is probably not perfect due to possibly mixed edge shape and dangling bonds[13]. At high bias ($V_{ds}$=1V), $\lambda_{edge}$ is longer than low bias, in this case, it is possible that optical phonon scattering limits the total mfp, with $\lambda_{op}$~10nm[15], similar to CNTs.

We next analyze how close the GNRFET operates to the ballistic performance limits by comparing experiments with theoretical modelling. The theoretical model computes the ballistic performance limits by assuming a single ballistic channel and ideal contacts (sufficiently negative SBs)[16]. We found that the $L$~236 nm device in Fig. 2a and b delivered about 21% of the ballistic current at $V_{ds}$=1V, and about 4.5% of the ballistic current at low $V_{ds}$<0.1V. The highest high bias ballisticity in our studied devices is ~38%. The ballisticity at low drain bias is consistent with the short edge elastic scattering mfp, but the large ballisticity at high drain biases is surprising, especially considering that optical phonon/zone boundary phonon (OP/ZBP) emission, which has a mfp of ~10nm, exists at high drain biases. The reasons could be similar to the small direct effect of OP scattering on the current in CNTFETs[17]. Because the OP/ZBP energy is high (~0.2eV), a carrier backscattered by emitting an OP/ZBP does not have enough energy to overcome the barrier near the source end of the channel, and return back to the source. Any subsequent edge scattering after OP/ZBP emission has a small direct effect on the DC current because edge scattering is elastic and does not change the carrier energy. Such a carrier rattles around in the channel and finally diffuses out of the drain. At high drain biases, therefore, only elastic scattering near the beginning of the channel matters and the rest of the channel essentially operates as a carrier absorber.

Compared to the earlier works on GNR of 20 nm width[18], the devices in current work show $10^5$ higher $I_{on}/I_{off}$ ratio at room temperature, ~20 times higher on current density (at $V_{ds}$=1V) and ~100 times higher transconductance per μm, due to larger band-gaps, high GNR quality with better edge smoothness[13], thin gate oxide and short GNR channel. At the same carrier concentration (e.g. $V_g$=-0.67V, corresponding to -20V on 300nm SiO$_2$) and $V_{ds}$=1V, our wide GNR devices deliver higher current density (~2000-3000μA/μm) than previously reported bilayer GNR with similar width (~50μA/μm)[12]. After correction for ~10 times channel lengths difference, our current levels are still a few times higher, indicating good GNR quality.

To further access the performance of our GNRFETs, we compared with CNTFETs. We fabricated Pd contacted CNTFETs on 10nm SiO$_2$ with similar channel lengths. The performances of our CNTFETs ($I_{on}$ and $I_{on}/I_{off}$ ratio) are very similar to previously published results[19]. We compared the on current density with different diameter CNTs at the same power supply voltage $V_{dd}$=$V_{ds}$=0.5V and $I_{on}/I_{off}$ ratio[20]. We used $V_{gs}$(on)-$V_{gs}$(off)=2V, equivalent to a 10nm gate dielectrics with dielectric constant ε≈4×3.9=15.6. In Fig. 4b, we plotted two representative GNRFETs with $w$~3nm, $L$~100nm and $w$~2nm, $L$~236nm, and compared them with $d$~1.6nm, 1.3nm and 1.1nm CNTFETs with similar channel lengths. Both GNRs have on current density ~2000μA/μm. The $d$~1.6nm CNTs outperform GNRs in terms of on current density (>3000μA/μm) but exhibit high off state leakage and a maximum $I_{on}/I_{off}$ ratio <$10^3$. For $d$~1.3 nm CNTs, they outperform GNRs in current density at the same

$I_{on}/I_{off}$ ratio (Fig. 4b). The $d$~1.1nm CNTs on the other hand, deliver much lower current density than GNRs at the same $I_{on}/I_{off}$ ratio, probably due to large positive SB, short AP mfp[21] and defects[22].

Our sub-10nm GNRFETs afford all-semiconducting nano-scale transistors that are comparable in performance to small diameter carbon nanotube devices. GNRs are possible candidates for future nano-electronics. Future work should focus on elucidating the atomic structures of the edges of our GNRs and correlate with the performances of GNRFETs. The integration of ultra thin high-κ dielectrics[23] and more aggressive channel length scaling is also needed to achieve better electrostatics, higher $I_{on}$ and ideal subthreshold slope.

This work was supported in part by MARCO MSD Focus Center and Intel.

**FIGURE CAPTIONS**

**Figure 1.** GNRFET device images and raman spectrum. **(a)** Schematics of GNRFETs on 10nm $SiO_2$ with Pd S/D. $P^{++}$ Si is used as backgate. **(b)** AFM image of a $w$~2±0.5nm, $L$~236nm GNRFET. Scale bar is 100nm. **(c)** AFM image of a $w$~60±5nm, $L$~190nm wide GNR device. Scale bar is 100nm.

**Figure 2.** Transistor performance of GNRFETs. **(a)** Transfer characteristics (current vs. gate voltage $I_{ds}$-$V_{gs}$) under various $V_{ds}$ for the device shown in Fig. 1b. $I_{on}/I_{off}$ ratio of >$10^6$ is achieved at room temperature. **(b)** Output characteristics ($I_{ds}$-$V_{ds}$) under various $V_{gs}$ for the device shown in Fig. 1b. On current density is ~2000μA/μm in this device. **(c)** Transfer and **(d)** output characteristics of the device shown in Fig. 1c.

**Figure 3.** Three channel lengths GNRFET. **(a)** AFM image of a typical $w$~2.5±1nm GNRFETs with three channel lengths. White arrows are pointing to the channels. $L$~110nm, 216nm and 470nm for the lower, middle and upper segments, respectively. Scale bar is 200nm. **(b)** Output characteristics (symbols) and simulations (lines) for the upper segment ($L$~470nm) of the device in **(a)**. From bottom, $V_{gs}$ is from -2V to 0.4V, with 0.4V/step. **(c)** Measured low bias on-state resistance (symbols) and linear fit (line) of the three segments in **(a)**. The extrapolated $R_c \approx 60$kΩ.

**Figure 4.** GNRFETs and CNTFETs performance comparison. **(a)** Mobility vs. $w$ for multi-channel GNRFETs. All data here were obtained from multi-probe measurements of single ribbons to exclude contact resistance **(b)** Current density (current normalized by $w$ for GNRs, $2d$ for CNTs) as a function of $I_{on}/I_{off}$ under $V_{dd}$=$V_{ds}$=0.5V and $V_{gs}$(on)-$V_{gs}$(off)=2V. Red symbol: $w$~3nm $L$~100nm GNR; blue symbol: $w$~2nm $L$~236nm GNR; black dashed line: $d$~1.6nm $L$~102nm CNT; black solid line: $d$~1.6nm $L$~254nm CNT; grey dashed line: $d$~1.3nm $L$~110nm CNT; grey solid line: $d$~1.1nm $L$~254nm CNT.

Figure 1.

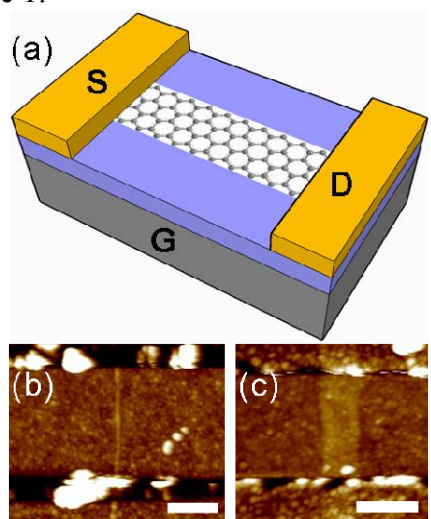

Figure 2.

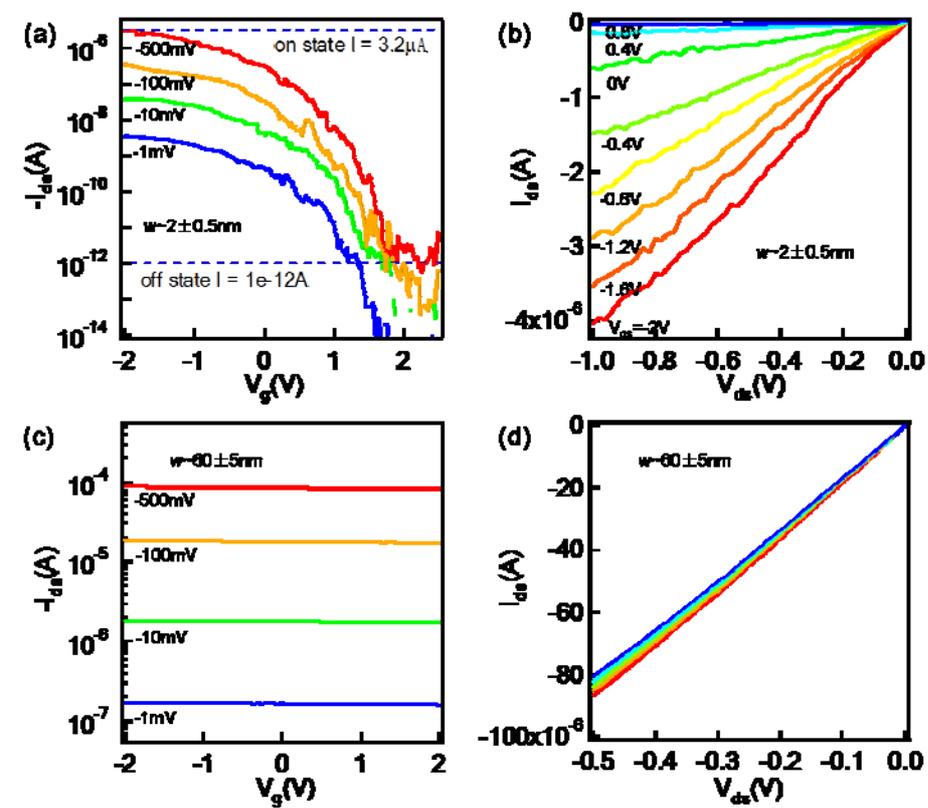

Figure 3.

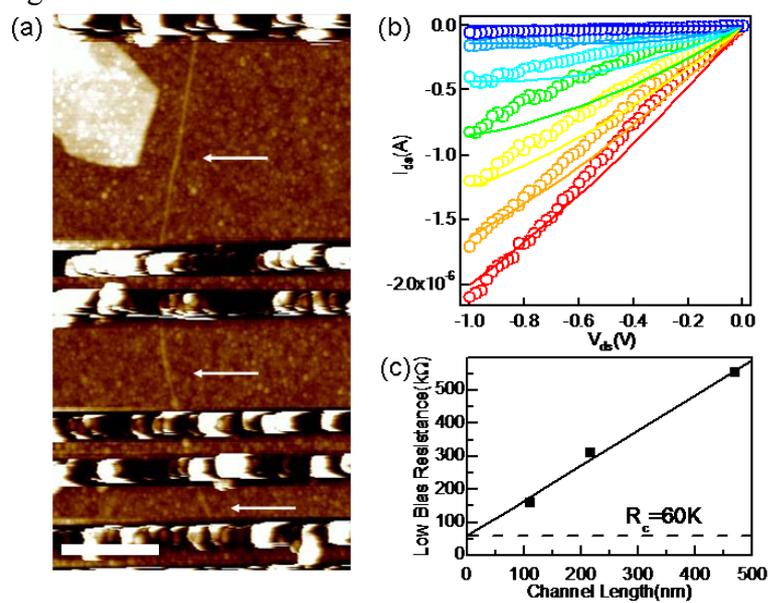

Figure 4.

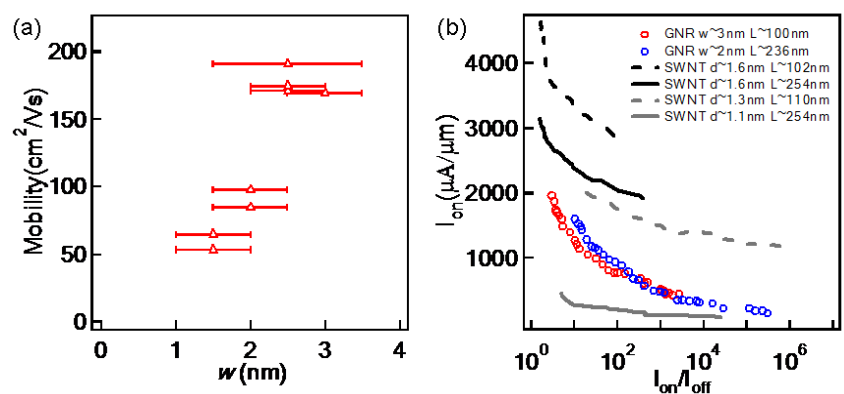

**Supplementary Information**

**Fabrication process of GNRFETs**

Following the synthesis steps described in Ref. 9 in main text, we obtained the graphene suspension in PmPV/DCE solution. We soaked the 10nm $SiO_2/p^{++}$ Si substrate with pre-patterned metal markers (2nm Ti/20nm Au) in the solution for ~20mins, rinsed with isopropanol and blew dry with argon. Then the chip was calcined in air at 350ºC for ~10mins and annealed in vacuum at 600ºC for ~10mins to further clean the surface. We used tapping mode AFM to find GNRs around the pre-patterned markers and recorded the location. Next we used electron beam lithography to pattern the S/D of the devices. 20nm Pd was then thermally evaporated as contact metal followed by a standard lift-off process. Finally, we annealed the device in argon at 200ºC for ~15mins to improve the contact.

**Confocal surface enhanced raman spectroscopy (SERS) study of GNRs.**

We carried out confocal SERS[1] mapping on GNR devices using a 633nm HeNe laser. After we took the electrical data of our GNRFETs, we evaporated ~5nm Ag on the chip and studied their raman spectra. We used a Renishaw inVia Raman microscope with an 80X objective lens operating in confocal mode. The collection area was ~1μm × 1μm. We used 633nm HeNe laser as excitation as it was more resonant with GNRs than 785nm laser. Low laser power of ~1mW was used in all the mapping experiments to prevent damage and thermal effects. In order to get good signal to noise ratio, the integration time in each spot was ~50 seconds. We usually mapped ~5μm × 5μm area centred on the GNR, and recorded D, G (~1200 – 1800$cm^{-1}$) and RMB (~100 – 300$cm^{-1}$) bands at each spot. The spectrum shown below was taken at the spot corresponding to the position of GNR by AFM image.

Fig. S1 shows the G (~1600$cm^{-1}$) and D (~1300$cm^{-1}$) band raman spectrum of the device in Fig. 1b of main text (also in the inset). We also measured the low frequency band (100 ~ 300$cm^{-1}$) of our GNR devices, but never observed any RBM peak in that region, which appeared to be the main difference between GNRs and carbon nanotubes (CNTs)[2]. In the G band, some GNRs showed triplet structure, with the main peak located ~1580-1590$cm^{-1}$, a hump like peak in lower energy, whose frequency depends on the widths of the GNRs, and a weak higher frequency peak ~1610$cm^{-1}$. The splitting of G peak is due to phonon confinement in the width direction, as in the case of CNTs[2-4]. Interestingly, for this particular GNR, the splitting between the main peak and lower frequency peak is ~55$cm^{-1}$, close to that of *d*~0.9nm (~3nm circumference, close to GNR width) semiconducting CNTs[5]. The D band intensity of some of the GNRs is high likely due to the edges, which is the main source of D band in large graphene[6-8]. The D and G bands of GNRs are broader than those in CNTs and bulk graphene. The relatively high D peak intensity and broadening of D and G peaks suggest the presence of $sp^2$ bond disorders[6,9], most likely on edges. Detailed raman study of these narrow GNRs is submitted separately[2].

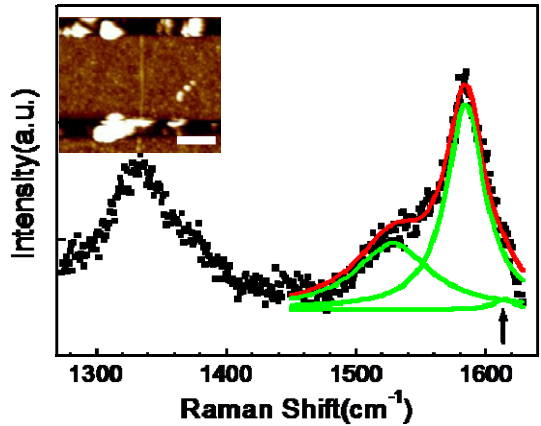

Figure S1. G and D bands raman spectrum (symbol) of the device in Figure 1b in main text. Red line shows the fit of G band using three Lorentzians (green lines). The peak positions (full widths at half maximum) are at 1529cm$^{-1}$ (70cm$^{-1}$), 1584cm$^{-1}$ (36cm$^{-1}$) and 1614cm$^{-1}$ (20cm$^{-1}$, indicated by an arrow). Inset is the AFM image of this device.

**Calculation of gate capacitance**

We used Fast Field Solvers (available at http://www.fastfieldsolvers.com) for three dimensional simulation of $C_{gs}$. The simulated structure included a large back plane as backgate, a dielectric layer with same lateral dimension of backgate, 10nm thickness and $\varepsilon_0$ =3.9, a graphene layer with experimental dimension lying ~0.5nm above the dielectric layer (note that the 0.5nm separation has little effect in calculated $C_g$) and two metal fingers with experimental dimension to represent contacts. In order to get precise result, rather fine grids are used in simulation (~1nm for GNR).

**Modelling edge scattering mfp in GNRs**

The first conduction subband *E-k* relation of a semiconducting GNR can be approximately expressed as,

$$E(k) = \frac{3ta_0}{2}\sqrt{k_\parallel^2 + k_\perp^2},$$

where *t* is the nearest neighbour hopping parameter, $a_0$ is the C-C bonding length, $k_\parallel$ is the component of the wave vector *k* in transport direction, and $k_\perp$ is the component in the confinement direction. The *E-k* relation yields a half band gap of $\Delta = 3ta_0|k_\perp|/2$ and a carrier kinetic energy of $E_k = E(k) - \Delta$. Under the semiclassical description, a distance of $wk_\parallel/k_\perp$ is travelled along the transport direction between two turns in the zigzag path of the carrier. If the backscattering probability is assumed to be *P*, which depends on the quality of the edges, the edge backscattering mfp is $\lambda_{edge} = k_\parallel w/(k_\perp P)$.

By substituting $E_k$ and $\Delta$ into this equation, we obtain $\lambda_{edge} = \frac{w}{P}\sqrt{\left(1+\frac{E_k}{\Delta}\right)^2 - 1}$. This expression indicates that the increase of the GNR width or carrier kinetic energy, and the improvement of the GNR edge quality can lead to an increase of the edge backscattering mfp. The above derivation were based on the simplified assumptions of an approximate *E-k* relation and a semiclassical approach, but the quantum simulation based on an atomistic simulation of a GNR with edge roughness using the non-equilibrium Green's function (NEGF) formalism with a numerical extraction of the mfp indicates that the above equation provides a simple and valid estimation of the edge backscattering mfp.

**Fabrication and performance of CNTFETs.**

The CNTs were grown by patterned CVD on 10nm $SiO_2$/$p^{++}$ Si substrate. Ebeam lithography was used to write S/D followed by Pd evaporation and lift-off. Devices were then annealed 200ºC in argon for 15mins to improve contact and probed to get electrical data.

Fig. S2 shows the transfer characteristics and AFM images of the devices used in Fig. 4b of main text.

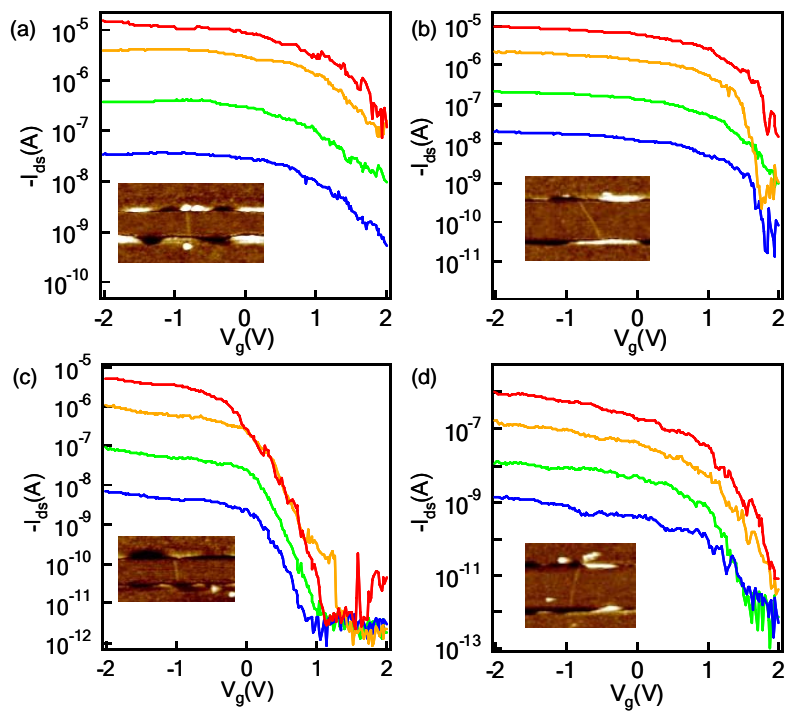

Figure S2. Transfer characteristics of CNTFETs. Drain bias $V_{ds}$=-1mV, -10mV, -100mV and -500mV for blue, green, orange and red curves, respectively. The insets are corresponding AFM images. **(a)** *d*~1.6nm *L*~102nm. **(b)** *d*~1.6nm *L*~254nm. **(a)**

*d*~1.3 nm *L*~110nm. **(a)** *d*~1.1nm *L*~254nm.